\documentclass[manuscript]{aastex}
\usepackage{color}
\usepackage{epstopdf}

\newcommand{\Jybm}{Jy~beam$^{-1}$}
\newcommand{\mJybm}{mJy~beam$^{-1}$}
\newcommand{\kms}{km~s$^{-1}$}
\newcommand{\vlsr}{$v_{\rm LSR}$ }

\shorttitle{The magnetic field in the NGC 2024 FIR 5}
\shortauthors{Alves et al.}

\begin{document}


\title{The magnetic field in the NGC 2024 FIR 5 dense core}


\author{Felipe de O. Alves and Josep M. Girart}
\affil{Institut de Ci\`encies de l'Espai (IEEC--CSIC), Campus UAB, Facultat de 
Ci\`encies, Torre C5 -- parell 2, 08193, Bellaterra, Catalunya, Spain}
\email{[oliveira;girart]@ice.cat}

\author{Shih-Ping Lai}
\affil{Institute of Astronomy and Department of Physics, National Tsing Hua 
University, Hsinchu 30043, Taiwan}
\email{slai@phys.nthu.edu.tw}

\author{Ramprasad Rao}
\affil{Submillimeter Array, Institute of Astronomy and Astrophysics, Academia 
Sinica, 645 N Aohoku Pl, HI 9672}
\email{rrao@sma.hawaii.edu}

\and

\author{Qizhou Zhang}
\affil{Harvard-Smithsonian Center for Astrophysics, 60 Garden Street, 
Cambridge, MA 02138, USA}
\email{qzhang@cfa.harvard.edu}


\begin{abstract}
We used the Submillimeter Array (SMA) to observe the thermal polarized dust emission from the protostellar 
source NGC 2024 FIR 5. The polarized emission outlines a partial hourglass morphology for the plane-of-
sky component of the core magnetic field. Our data are consistent with previous BIMA maps,  
and the overall magnetic field geometries obtained with both instruments are similar. We resolve the 
main core into two components, FIR 5A and FIR 5B. A possible explanation for the asymmetrical 
field lies in depolarization effects due to the lack of
internal heating from FIR 5B source, which may be in a prestellar evolutionary state. 
The field strength was 
estimated to be 2.2 mG, in agreement with previous BIMA data. We discuss the influence 
of a nearby H{\sc ii} region over the field lines at scales of $\sim 0.01$ pc. Although the 
hot component is probably compressing the molecular gas where the dust core is embedded, 
it is unlikely that the radiation pressure exceeds the magnetic tension.
Finally, a complex outflow morphology is observed in CO (3 $\rightarrow$ 2) maps. Unlike previous maps, 
several features associated with dust condensations other than FIR 5 are detected. 
\end{abstract}


\section{Introduction}
\label{intro}

Understanding the evolution of molecular clouds and protostellar cores is one of 
the outstanding concerns of modern astrophysics. Particularly, efforts are
concentrated  in determining which physical agents are mainly responsible for
controlling the dynamical  properties of the dense cores. It is widely accepted by the
astronomical community that magnetic fields must be taken into account in
evolutionary models of collapsing protostellar cores \citep{Shu99}. Although
some  theories claim that turbulent supersonic flows drives star formation in the
interstellar medium  \citep{Elmegreen04,Maclow04}, others demonstrate
that the ambipolar diffusion collapse theory  reproduces properly observed
molecular cloud lifetimes and star formation timescales 
\citep{Tassis04,Mouschovias06}.

One way to resolve these open issues in star forming theory is to increase the
number of high-quality observations which resolve the core collapse structure. 
Sampling several protostellar cores with
distinct physical properties can provide better constraints to improve simulations. 
Particularly, the number of observations of magnetic fields
in molecular clouds and dense cores has been increasing rapidly with the advent
of new instruments with high sensitivity. In polarimetry, it is globally accepted that 
non-spherical dust  grains are aligned perpendicular to field lines \citep{Davis51} producing linearly polarized
thermal  continuum emission \citep{Hildebrand88}. Which mechanism mainly
contributes to the alignment of interstellar dust grains is still a matter of
debate \citep{Lazarian07}. However, very recently  \citet{Hoang08,Hoang09} have
successfully modeled the polarization by radiative torques propelled by 
anisotropic radiation fluxes. Those torques act to align spinning non-spherical dust grains with their 
largest moment of inertia axis parallel to the field lines. The polarized flux is usually only a small fraction
of the total intensity (usually a few  percent) and for this reason the study of the magnetic
field is highly limited by the instrumental sensitivity. 

Cold dust emits mainly at far-IR and submm wavelengths. In the submm regime, the emission 
is optically thin and, therefore, it is not
affected by scattering or absorption. For this reason, the SMA
has been extensively used to study thermal emission from  dust and cool gas. Several
authors reported polarization observations of different classes of protostellar cores.
The textbook case is the low mass young
stellar system NGC 1333 IRAS 4A  \citep{Girart06}. The supercritical state is
reflected in the SMA polarization maps which indicates a clear  hourglass
morphology for the  plane-of-sky magnetic field component in a physical scale of 300-1000
AU. This remarkable result not only was predicted by theories of collapse of
magnetized clouds \citep{Fiedler93,Galli93} but is commonly used to test models
of low-mass collapsing cores \citep{Shu06,Goncalves08,Rao09}. In the regime
of high mass protostars, recent investigations performed with the SMA also provided
observational constraints  on the physics involved during the core collapse
stage. Two recent works exemplify distinct  magnetic field
features within this class of objects. The polarimetric properties of G5.89--0.39
are consistent with a complex, less ordered field likely disturbed by an
ionization front \citep{Tang09}, while the hourglass morphology expected for
magnetically supported regimes was observed at large physical scales ($ \sim$
10$^{4}$  AU) for the protostellar core G31.41+03 \citep{Girart09}. Despite
the distinct energy balance  and timescales of the two regimes (low and high
mass), both results imply that the magnetic support must not be ignored in the
models.

\begin{figure}
\epsscale{0.60}
\plotone{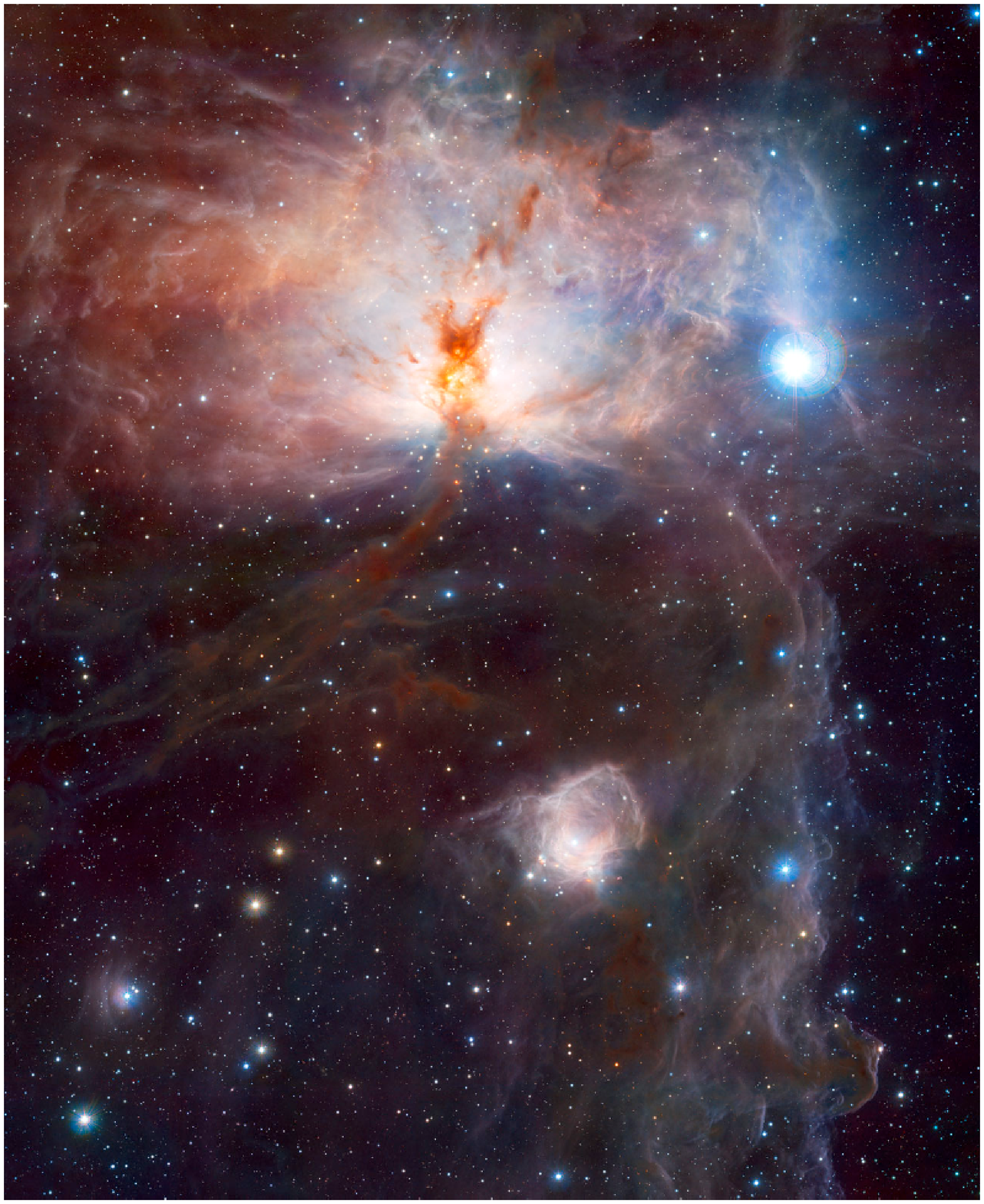}
\caption{VISTA image of the Flame Nebula (NGC 2024).
     The obscuring dust lane that exists foreground to the bright H{\sc ii} emission
     is seen almost transparent in this near infra-red image. The glow of 
     NGC~2023 and the Horsehead Nebula are seen in the lower portion of the 
     image.} 
\label{ngc2024}
\end{figure}

NGC 2024 is the most active star forming region in the Orion B giant molecular
cloud. The gas structure in this region has an ionized component surrounded by a
background dense molecular ridge and a foreground dust and  molecular component
visible in the optical images as a dark lane. Recently, the new ESO telescope 
VISTA (Visible and Infrared Survey Telescope for Astronomy) released a high 
sensitivity near-infrared image of  NGC 2024 (Figure \ref{ngc2024}). In this 
large scale view, the foreground dust lane which optically obscures 
the H{\sc ii} region is almost transparent. Scattered light produced by 
the ionization front is seen as bright emission in the top of the image, and 
a cluster of hot young stars is revealed. \citet{Kandori07} used
near-infrared polarimetry  to study the scattered light from the H{\sc ii}
region. Several reflection nebulae associated with young stellar  objects (YSO)
were discovered in their polarization maps. The overall centro-symmetric pattern
suggests that the  ionizing source is IRS 2b, a massive star located $5''$
north-west of IRS 2, in agreement with a previous near-infrared photometric study carried
out by \citet{Bik03}. The submillimeter continuum emission arising from the dense
molecular  ridge was first observed by \citet{Mezger88,Mezger92}. Several far
infrared cores (so the acronym ``FIR'') at  distinct evolutionary stages were
identified and catalogued in a North-South (NS) distribution. 
In fact, the chain of FIR cores could have been generated by the interaction 
between the nearby H{\sc ii} region and the surrounding molecular cloud. 
\citet{Fukuda00} performed numerical 
simulations of sequential star formation trigged by an expanding H{\sc ii} region near a filamentary
cloud. In their models, isothermal expansion and magnetohydrodynamic effects are 
considered. Their simulations preview that a chain of cores is formed from this interaction, each pair of
cores belonging to a distinct generation though. Comparison between model and the dynamical 
parameters observed in NGC 2024 are quite good. In particular, they state that FIR 4 and FIR 5 belong to 
the first generation of cores, what is confirmed by the observed dynamical ages of their outflows.
In this paper, 
we center our discussion on FIR 5, the brightest and most evolved of them,  with a strong 
and collimated unipolar outflow \citep{Richer92}. Continuum observations at 3 mm 
performed by  \citet{Wie97} suggest that FIR 5 is a double core embedded in an envelope. 
However, higher angular resolution  observations from \citet{Lai02} (LCGR02 from 
now on) resolved the dust emission in one strong component surrounded by several 
weaker components in a radius of a few arcseconds. 

Several authors have conducted polarimetric investigations toward FIR 5. 
\citet{Crutcher99} used the Very Large Array (VLA) to carry out Zeeman
observations of OH and H{\sc i} absorption lines in order to trace the
line-of-sight (LOS) component of the magnetic field.  These authors found a LOS
field gradient of $\sim 100~\mu$G across the northeast-southwest direction. Dust 
emission polarization in the surroundings of FIR 5 was mapped in 100~$\mu$m by
\citet{Hildebrand95} and  \citet{Dotson00} with the Kuiper Airborne Observatory.
At longer wavelengths, \citet{Matthews02} used the SCUBA polarimeter 
to observe the 850~$\mu$m emission with at the James Clerk Maxwell 
Telescope (JCMT)  and obtained polarization
patterns consistent with those derived with the 100~$\mu$m data. 
Their single-dish
dust  polarization maps trace a relatively ordered magnetic field along the ridge of
emission containing FIR 4 and FIR 5 (Fig. 4 in their paper). Based on the spatial
coverage of their observations, this ordered field must extend over a distance of 
at least $\sim 0.5$~pc (for the  assumed distance of 415~pc to the Orion B 
cloud). \citet{Matthews02} modeled this field using a helical-field geometry threading 
a curved filament, since this configuration suited fairly to a 2-dimensional projection accordingly to the 
SCUBA maps. However, they found that this geometry is not consistent with the LOS Zeeman data
of \citet{Crutcher99} because no reversed fields are seen at both sides of the chain of cores. 
Instead, those authors offer another interpretation based on a compression zone created due to the 
expansion of the foreground ionization front. In this scenario, the
magnetic field lines are stretched  around the ridge of dense cores in a
physical morphology consistent with LOS field gradient observed in the Zeeman data of \citet{Crutcher99}. 

Concerning the local field associated with FIR 5, the work of Lai et al. (2002, hereafter LCGR02)
has the best resolution for the dust continuum emission so far.
These authors used the 
Berkeley-Illinois-Maryland-Association (BIMA array) interferometer and obtained an
angular resolution of $2\farcs4 \times 1\farcs4$. The polarized flux of the BIMA maps
extends in a N-S direction, perpendicular to the putative  protostellar
disk. The corresponding field lines were fitted with a geometric model consisting
of a set of concentric parabolas, indicating that the polarized flux trace a
partial hourglass morphology for the magnetic field. In this paper, we report
SMA dust continuum polarization toward FIR 5. The higher 
sensitivity of this instrument provides new information on the detailed field morphology of the FIR 5
core.

\section{Observations and Data Reduction}
\label{sec_obs}

The high angular resolution of the SMA allows us to trace the thermal 
emission of dust grains at physical scales 
of few hundred astronomical units\footnote{The Submillimeter Array is a joint project between the 
Smithsonian Astrophysical Observatory and the Academia Sinica Institute of Astronomy and
Astrophysics, and is funded by the Smithsonian Institution and the Academia
Sinica.} (for objects in the Orion 
molecular cloud complex) and, therefore, is able to spatially resolve compact 
dust  cores. A detailed description of SMA is given in \cite{Ho04}.  
The observations were carried out in 2007 November 24 and December 19 with 
the SMA in its compact configuration. The number of antennas available for 
the observations were 7 and 6, respectively. The atmospheric opacity at 225 GHz 
was  0.11 and 0.07 for the first and second day, respectively (values measured 
by the Caltech Submillimeter Observatory tau meter). Observations were
done in the 345 GHz atmospheric window, what corresponds to a wavelength 
of 870 $\mu$m. The SMA receivers operate in two 
sidebands separated by $\sim 10$~GHz. The central observed frequencies for 
the lower and upper side bands were 336.5 GHz and 346.5 GHz, respectively. 
The SMA correlator had a bandwidth of 1.9 GHz (for each sideband) divided in
24 ``chunks'' of 128 channels each. In total, the full-band spectrum contains 3072 channels 
for each sideband and a spectral resolution of 0.62 MHz, which corresponds to a 
velocity resolution of 0.7~km~s$^{-1}$. SMA receivers are single linearly 
polarized. By using a quarter-wave plate  in front of each receiver, 
the incoming radiation is converted into circular polarization (L, R).
The SMA correlator combines the signal into circular polarization vectors: 
RR, LL, RL, LR. In order to obtain the full four Stokes parameters for all the
baselines, the visibilities have to be averaged on a time scale of 5~minutes.
A description of the SMA polarimeter
and the discussion of the methodology (both hardware and software aspects) are
available in \citet{Marrone06} and \citet{Marrone08}.

The phase center 
($\alpha_{2000} = 05^{\mathrm{h}} 41^{\mathrm{m}} 44\fs3$,
$\delta_{2000} =  -01\degr55' 40\farcs8$) 
was set according to the peak of emission
obtained for FIR 5 in LCGR02. 
Uranus and Titan were  observed as flux
calibrators in both tracks. The resulting visibility function for each
calibrator is consistent with the  expected flux estimated by the SMA Planetary
Visibility Function Calculator during the observing runs. The quasar J0528+134 
was used as the gain calibrator.  The quasar 3c454.3 was used for bandpass and
polarization calibration.  The first track had a much better parallactic angle
coverage for 3c454.3 than the second track, thus the  3c454.3 data from the first
track were used to solve for the instrumental polarization or ``leakages''. 
The minimum and maximum UV distance 
for both tracks was 16 and 88 k$\lambda$, respectively.
Antenna 3 was used only in the second track, so no leakage solution 
could be derived. Thus, antenna 3 was not used to obtain Stokes Q 
and U maps.  After the calibration steps, data from upper and lower sidebands for each 
track were synthesized into a single data set.  Calibrated visibilities for 
each track were combined into a final data set.  Removal of continuum 
contamination from the line data set was done. The main contribution arose 
from the CO ($3\rightarrow2$) transition at the chunk \# 4 of the upper sideband 
($\sim 345.76$  GHz)\footnote{Since our interest in the line data set concerns only
Stokes I emission, antenna 3 is unflagged in the deconvolved CO ($3\rightarrow2$)
maps.}. 

All the calibration and reduction steps were done with MIRIAD configured for SMA
data \citep{Wright93}. The science target was strong enough and
self-calibration was performed in order to increase the signal-to-noise ratio in
the final maps.  Imaging of the Stokes parameters I, Q and U was performed. 
Maps of polarized intensity ($I_{P}$), polarized fraction ($P$) and position of polarization
angles ($\theta$) were obtained by combining Q and U images in such a way that
$P = \frac{I_{P}}{\mathrm{I}} = \frac{\sqrt{Q^{2} + U^{2}}}{\mathrm{I}}$ and $\theta = 
\frac{1}{2} \mathrm{tan}^{-1}(\frac{U}{Q})$. The resulting synthesized 
beam of Stokes I maps has $2\farcs45 \times 1\farcs48$, with a position angle 
(PA) of $-39.8\degr$. Table \ref{table_cont} summarizes the technical parameters 
of continuum and line observations.  

\section{Results}

\subsection{Dust Continuum Emission}
\label{prop_dust}
 
\begin{figure}
   \centering
   \includegraphics[angle=0,width=11.5cm]{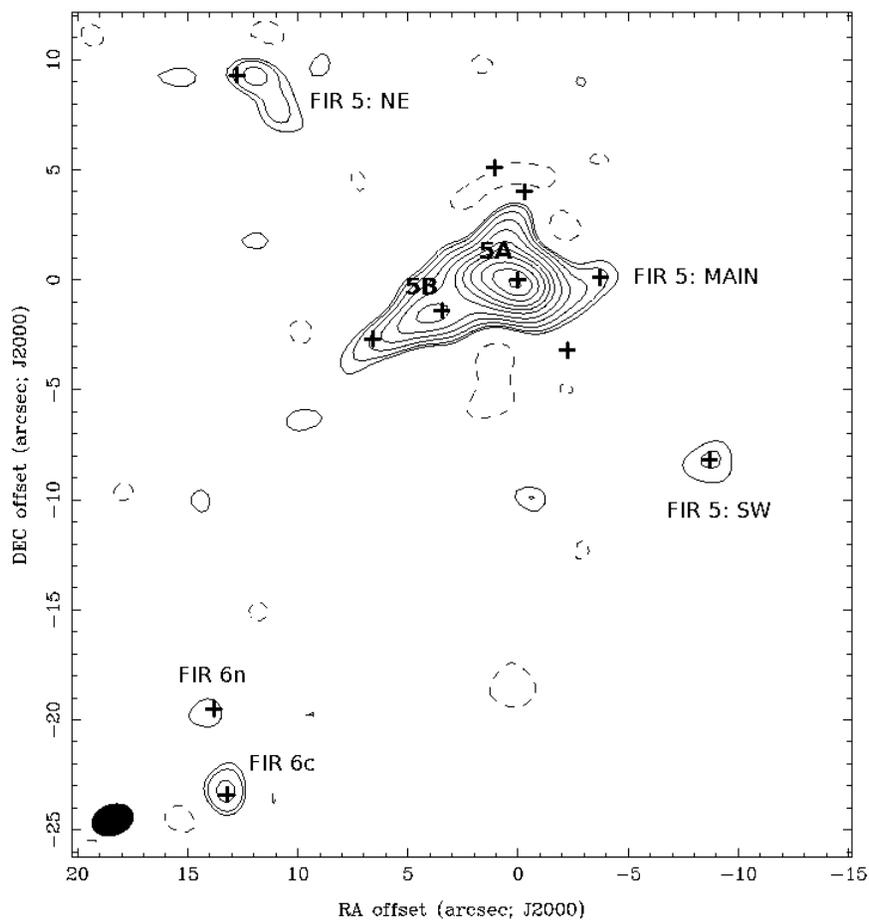}
      \caption{Dust continuum map of FIR 5 with quasi-uniform weighting 
      ($robust = -1$). Contours are drawn at $-3$, 3, 4, 6, 9, 13, 19, 25, 32, 
      42, 52, 62 $\sigma$ (1--$\sigma \simeq 18$~\mJybm). The half power 
      beam width (HPBW) of the synthesized beam is $1\farcs96\times1\farcs41$ 
      and the position angle is $-70.6\degr$. The crosses indicate the dust 
      continuum sources detected by \citet{Lai02} with BIMA.} 
      \label{fir5_highres}
   \end{figure}

Figure~\ref{fir5_highres} shows the contour map of the 878~$\mu$m dust 
emission in FIR 5 obtained with a quasi-uniform weighting (a robust of $-1$), 
which provides a better angular resolution of $1\farcs96 \times 
1\farcs41$. The overall submillimeter
emission resembles the 1.3~mm dust continuum maps obtained with BIMA by 
LCGR02, although the latter has a $rms$ a factor of 3 lower. 
Our observations (with shortest baseline equal to 16 k$\lambda$) allow us to only 
measure structures smaller than $\sim$ 5.7 arcseconds (see Equation A.5 of \citet{Palau10}). 
In LCGR02, they find extended emission at scales of 5-7 arcseconds with BIMA. 
The brightest emission arises around FIR 5: Main (following 
LCGR02 notation), which is resolved into two components, 5A and 5B. 
Those components
correspond to the double source detected in 3 mm by \citet{Wie97} and indentified as
FIR 5-w and FIR 5-e. However, not all the fainter sources observed in the FIR 5: Main 
core of LCGR02  have been detected with the SMA. FIR 5: Main appears more 
extended in the BIMA maps, attributable to the better
sampling of shorter baselines with the BIMA array. In particular, the N-S direction
contains emission of
the dust condensations LCGR2, LCGR3 and LCGR5 (according to LCGR02 nomenclature).
These sources are missing in our SMA maps probably due to the absence of antenna 3 in the
deconvolved maps (see section \ref{sec_obs}). Antennas 3 and 6 cover a short baseline in the
UV plane which is parallel to the U axis and close to $V = 0$ k$\lambda$. In equatorial coordinates, it 
corresponds to features parallel to the declination axis, and close to the phase center. Therefore, by
flagging antenna 3 we lost this flux component which should be produced by the missing sources.
In addition, different 
visibility sampling between the two aperture synthesis telescopes (BIMA
sampled shorter baselines) must also be considered.
We also 
detect several faint peaks (at the 4 and 7--$\sigma$ level; 1 $\sigma = 18$ \mJybm) also seen 
by LCGR02.

Tables~ \ref{fir5main_table} and \ref{condens} give the dust emission 
properties for the two condensations associated with FIR 5: Main and for 
the fainter dust condensations, respectively. The intensity peak and the 
position of the sources were derived using the Miriad task ``maxfit''.
For the two bright sources associated with FIR 5: Main, a two Gaussian fit
(using the AIPS's ``imfit'' task) was used to estimate the flux density 
of each component and its size. The two sources appear resolved but in 
different directions. Thus, source 5A has a full width half maximum (FWHM) 
size of $2\farcs8\times2\farcs4$ elongated close to the NE--SW axis 
($PA=67\arcdeg$), whereas source 5B has a deconvolved FWHM size of $3\farcs6\times2\farcs8$ 
but is elongated along the SE--NW direction ($PA=124\arcdeg$).

In order to estimate the column density and mass of the cores, we 
need to assume a value for the cores' temperatures. Different molecular line 
observations have established that the NGC 2024 cores are warm with 
temperatures between 40 and 85~K \citep{Ho93,Mangum99,Watanabe08, 
Empre09}. Here we adopt a temperature of 60~K.  We assume a dust opacity 
at 878~$\mu$m of 1.5~cm$^{2}$~g$^{-1}$, which approximately
is the expected value for dust grains with thin dust mantles at densities of 
$\sim 10^6$~cm$^{-3}$ \citep{Ossenkopf94}. Using the previous FWHM 
sizes derived  from the Gaussian fit, a beam-averaged column density of 
$\sim 4.7 \times 10^{23}$ and $2.2 \times10^{23}$ cm$^{-2}$
for sources 5A and 5B were derived, respectively. Similarly, masses for these two components
are 1.09 and 0.38~M$_{\odot}$, respectively.
The total mass of FIR 5:Main, 1.5~M$_{\odot}$, is consistent with the value 
derived by \citet{Chandler96}.

\subsection{Distribution of the polarized flux}
\label{distribution}

\begin{figure}
   \centering
   \includegraphics[angle=0,width=10.0cm]{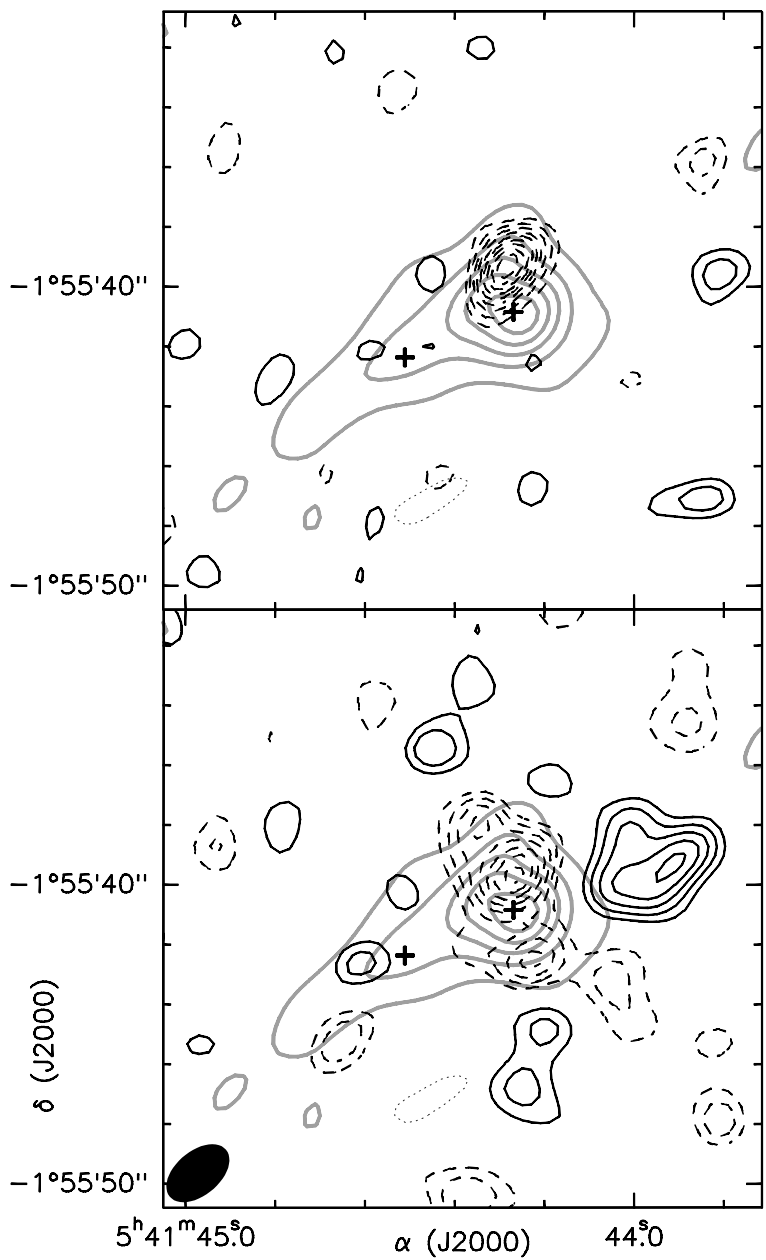}
      \caption{Maps of Stokes $Q$ ({\it top panel}) and $U$ ({\it bottom panel})
      emission. Dashed and solid thin  contours correspond to negative and positive
      polarized emission, repectively. The contours start at $-2$--$\sigma$ and 2--$\sigma$ 
      level with steps of 1--$\sigma$  (1--$\sigma = 5.3$~\mJybm). The absolute 
      $Q$ and $U$  peak fluxes are 0.056~\Jybm and 0.047~\Jybm, respectively. 
      The thick grey contours show the Stokes I emission.  Crosses indicate the position 
      of the two dust intensity peaks. The synthesized beam of the maps is shown 
      in the bottom left corner of the bottom panel.}
         \label{stokesqu}
   \end{figure}

For better sensitivity to the weak polarized emission, maps of Stokes 
I, Q and U were obtained with a robust weight of 0.50.  Figure~\ref{stokesqu} 
shows the Stokes Q and U emission, which have different distributions.
The Stokes Q emission arises from a negative compact spot about one 
arcsecond north of source A. The Stokes U is quite extended along
FIR5: Main, with the brightest emission around source 5A. Source 5B 
has only weak polarized emission at 3--$\sigma$ level. It is noteworthy that 
significant  positive Stokes U  emission appears west of source 5A without 
dust emission associated. However, this spot of polarized emission coincides 
with the BIMA continuum source FIR 5: LCGR 1 (catalogue of LCGR02). 
The non-detection by the SMA could occur because dust 
emission has been resolved out by 
the interferometer  (approximately 30\% of the flux is filtered out, 
see section \ref{prop_dust}). 
Thus, we tentatively associated this polarized spot to this source.
A cutoff of 2--$\sigma$ (1-- $\sigma \simeq 5$~\mJybm) in polarized intensity
($\sqrt{Q^{2} + U^{2}}$) is used to obtain the linear polarization emission and 
to derive the position angle in the plane of the sky of the polarization vectors.

The  polarization intensity and the polarization fraction in our maps achieve values as high as 
54 $\pm$ 6~\mJybm\ and 15\% $\pm$ 2\%,  respectively, at the northern portions of the core, where 
the polarized emission is brighter. Figure \ref{fir5pol} shows the dust 
continuum emission from the protostar overlaid with the dust polarization vectors. 
Using the position of the continuum peak 
as reference, three main components can be distinguished: a northern 
component, where the highest polarization degrees are obtained, a southwestern 
component  and an eastern component offset by $\approx 5''$ from the continuum 
peak.  This distribution is well represented in the histogram of polarization angles 
shown in Figure \ref{hist}.   There is a  change of roughly 90$\degr$ in the position 
angles of vectors 
associated with FIR 5A and the eastern vectors associated with FIR 5B.  Concerning only vectors 
associated with FIR 5A, position angles have a gradual rotation of approximately 40$\degr$ from 
north  to south. Table \ref{table} summarizes our polarization data. Note that 
the distribution of the polarized flux of the SMA data is remarkably consistent with 
the  BIMA maps of LCGR02. Although the structure of emission in both the BIMA and 
SMA data sets has the same overall pattern, the latter has a larger area of polarized flux. 
Compared to the JCMT maps of \citet{Matthews02}, the mean direction of our SMA 
polarization field is consistent with the lower resolution single-dish data, which
do not resolve the structure of FIR 5 and traces a larger physical scale associated with the
diffuse gas found at the core envelope.

\subsection{CO (3 $\rightarrow$ 2) emission}

Our SMA CO (3 $\rightarrow$ 2) maps reveal a very complex morphology
possibly related to multiple outflows. 
Figure \ref{CO32} shows the channel maps of the 
CO (3 $\rightarrow$ 2) emission with a velocity resolution of $\sim 2.1$ \kms. 
At blueshifted velocities the emission arises from an elongated but wiggling
structure in the East-West direction. This blueshifted component appears 
to be associated with FIR 6. 
The distribution of the molecular gas at the cloud systemic velocity 
(\vlsr$\simeq 10$~\kms)  is basically associated with the FIR 5 main core.
At redshifted velocities (\vlsr$\ga 14$~\kms) there are two main elongated 
features almost parallel extending in the North--South direction and observed 
over a wide range of velocities (up to \vlsr$\simeq 30$~\kms). One of these features is 
associated with the well-know outflow powered by 
FIR 5A \citep{Sanders85,Richer92,Chandler96} and the other one is located about $10''$ to the west 
and seems to arise from FIR 5-sw. These two lobes have 
their brightest emission located near their associated dust components (FIR 5A and sw).
The emission presents a clumpy morphology, with an average angular 
size of $\sim 5''$, corresponding to a physical size of 0.01 parsecs.
It is worth noting that the three possible CO high velocity features have no 
counterpart at the opposite flow velocities. Thus, the North-South redshifted lobes
have no blueshifted counterpart, and the East--West blueshifted lobe does not 
have a redshifted counterpart.

Figure~\ref{pvplot_co32} shows the Position-Velocity (PV) diagram centered in
FIR 5A with a PA = 0$\degr\!$.9 (along the brightest redshifted lobe).  
The outflow powered by FIR 5A has a wide distribution of velocities which prevails 
until $\sim 30$~\kms. An extended spatial distribution is observed up to $\sim 35^{\prime\prime}$ 
south of the source, although only low velocity components are observed at such distances.
No blue lobe is seen and only residual emission is measured in the northern 
counterpart. The blue component observed at the offset position of $-14^{\prime\prime}$ is part of
the outflow associated with FIR 6.
In section \ref{outflows} we provide a detailed discussion about the molecular 
distribution in this region.

\section{Discussion}

\subsection{Polarization properties}

\begin{figure}
   \centering
   \includegraphics[angle=-90,width=14cm]{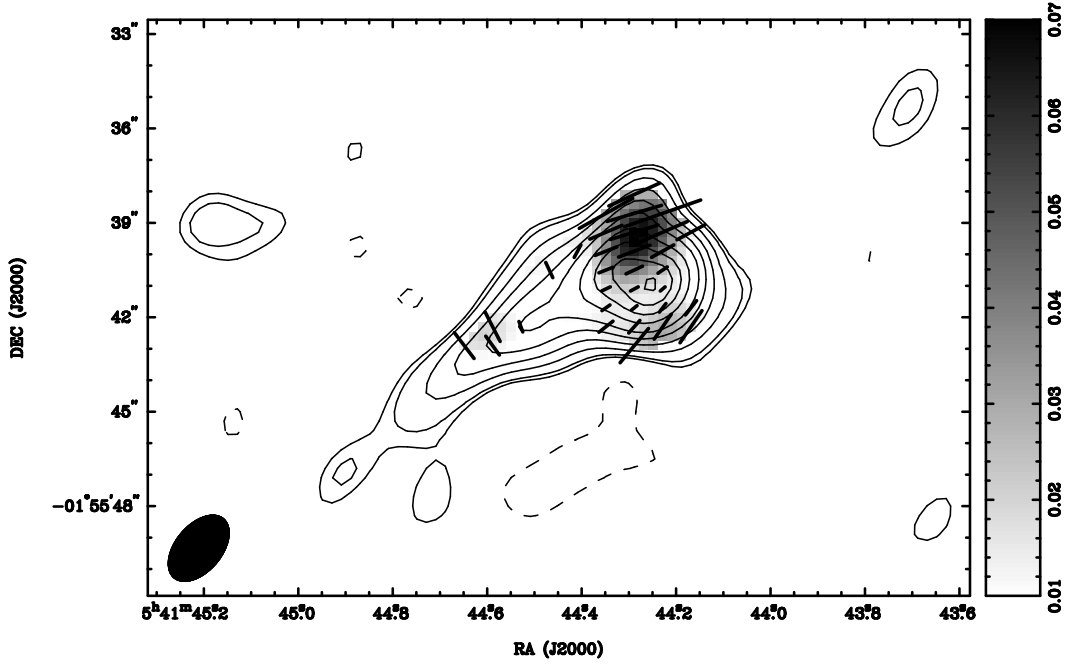}
      \caption{Contour map of the dust continuum emission overlapped with the 
      linear polarization vectors (black vectors) towards NGC 2024 FIR 5. Gray
      scale correspond to the polarized dust intensity.
      Contours levels are $-3$, 3, 4, 7, 11, 16, 22, 32, 42, 52, 62--$\times$  the 
      {\rm rms} noise of the dust emission ($\sim 19$~\mJybm). The length of 
      each segment is proportional to the degree of polarization (the length
      corresponding to a polarization level  of 10\% is indicated at the image 
      bottom) . The synthesized  beam is of $2\farcs45 \times 1\farcs48$ with
      a position angle of $-40\arcdeg$. Vectors are sampled as 2/3 of a beam.}
      \label{fir5pol}
   \end{figure}

\begin{figure}
   \centering
   \includegraphics[angle=-90,width=6.8cm]{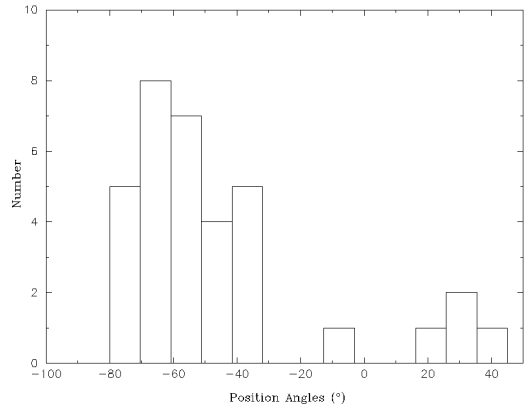}
      \caption{Histogram of position of polarization angles. The three polarized
components of FIR 5 polarization map are clearly seen in this plot.}
         \label{hist}
   \end{figure}

\begin{figure}
\centering
\includegraphics[width=16.5cm]{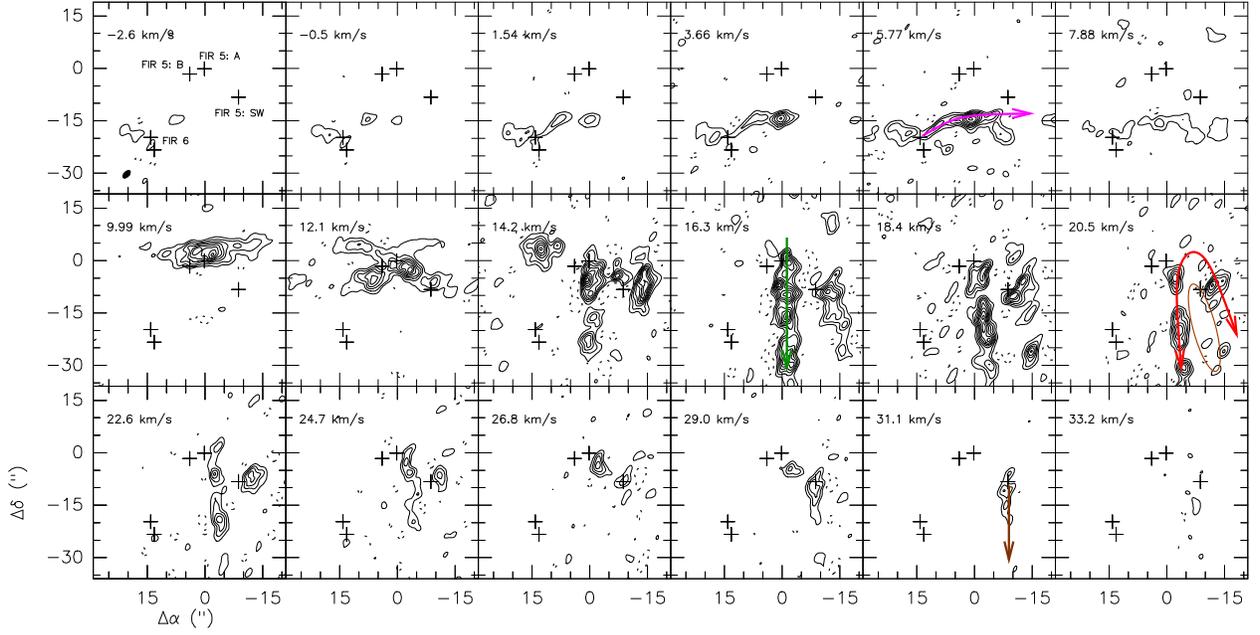}
\caption{Channel maps of the CO ($3 \rightarrow 2$) emission associated to
FIR 5 and FIR 6 dust cores. Contours levels are -4, 4, 8, 12, \dots to 36 $\times$ 
0.25~\mJybm (the rms noise of the map). The value of the V$_{LSR}$ is shown
in the top left corner of each panel. Source positions are indicated as crosses  
and labelled in the first panel. Magenta, green, red and brown arrows indicate the 
position of the FIR 6, FIR 5A, the precessing and the FIR 5-sw outflows respectively. 
An ellipse indicates the supposedly cavity produced by the high-velocity components of the
main FIR 5 outflow component.} 
\label{CO32}
\end{figure}

\begin{figure}
\centering
\includegraphics[angle=0,width=14cm]{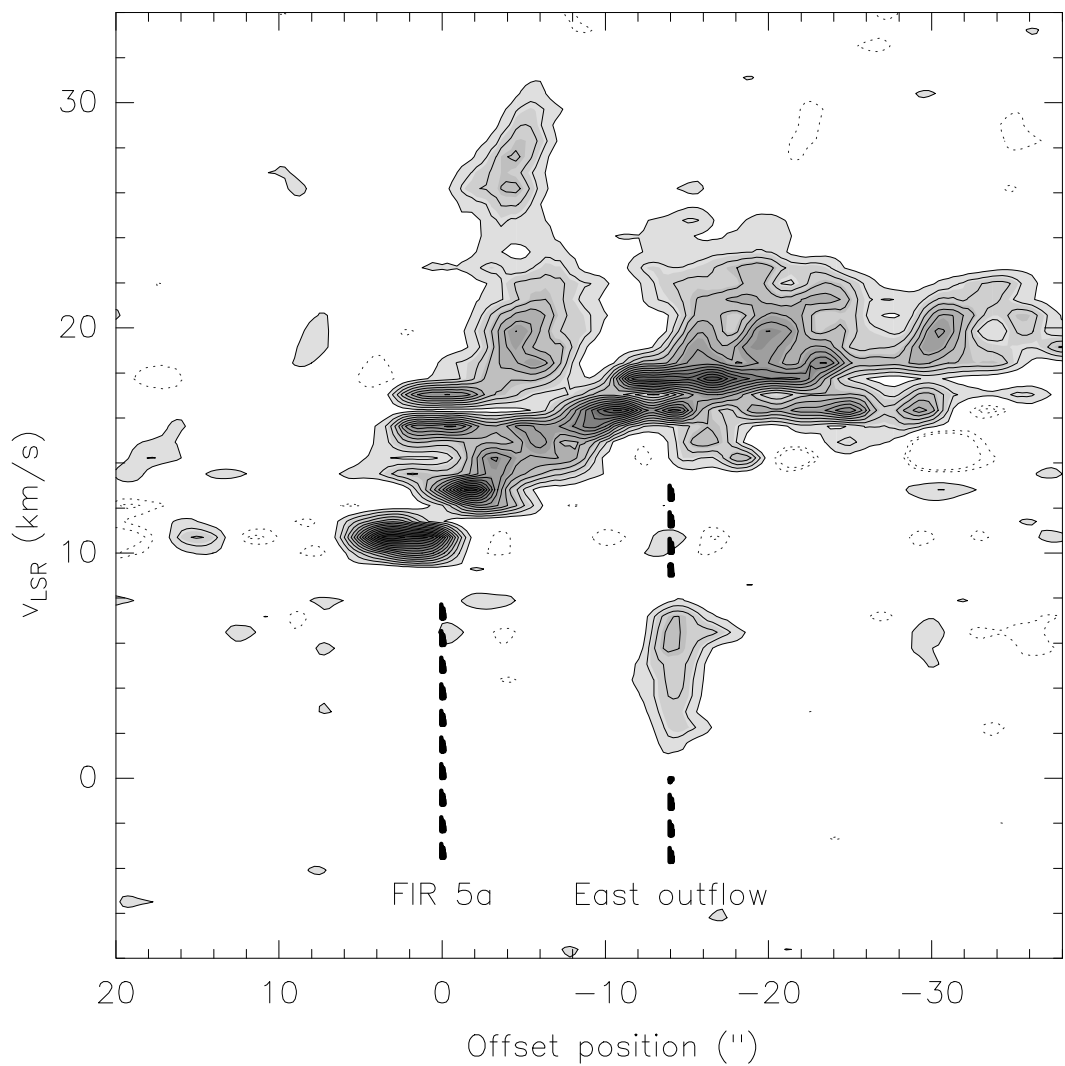}
\caption{Position-Velocity plot of the CO ($3 \rightarrow 2$) emission centered 
close to source FIR 5A (one arcsecond to the west) and along the North (positive
offsets) to South (negative offsets) direction. 
Contours levels are $-5$, $-3$, 3, and then steps of 3 times 0.3~Jy~beam$^{-1}$,
the $rms$ noise of the channel maps where the cut was obtained. The position of the 
driving source of the redshifted outflow, FIR 5A, is indicated with a dashed line.
The spatial overlap with the East ouftflow associated with FIR 6 is also
shown with a dashed line.}
\label{pvplot_co32}
\end{figure}

\begin{figure}
\centering
\epsscale{1.0}
\plotone{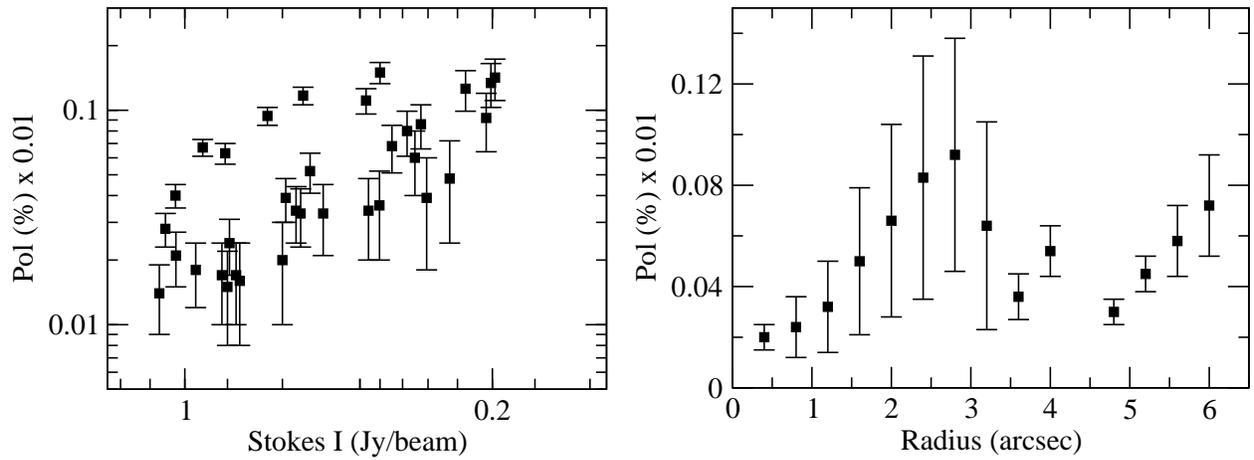}
\caption{Distribution of polarization toward NGC 2024 FIR 5. {\it Left
panel}: Polarization intensity  versus total intensity. {\it Right panel}: Polarization
intensity versus radius with respect to the  peak of Stokes I emission.}
\label{poldist}
\end{figure}

In this section, we focus our analysis on the northern and southwestern 
polarization features, which are the  brightest components and scientifically 
more interesting since a less uniform pattern is observed. At 2-$\sigma$ level, 
these two regions are connected and surround the peak of total intensity. 
From Figure \ref{fir5pol}, it can be noted that the peak of polarized and total
intensities are  approximately $2\farcs6$ apart. 
Figure \ref{poldist} shows the dependence of the polarization fraction with
Stokes I flux and with respect to the distance to source A. In both cases, there is a clear
depolarization toward the center, where the highest density portions of the core are located. 
The left panel of Figure \ref{poldist} suggests that 
the distribution of polarization with respect to the Stokes I emission seems to
be composed by two subsets: 
the highest polarization fraction data that has a slower growing curve and
corresponds to the northern component, and the subset with a linear 
dependence, which arises from the southwestern component. The right 
panel of Figure \ref{poldist} was produced by performing averaging over 
polarization data for concentric annuli of  $0\farcs4$ each. 

The depolarization effect is observed not only at the brightest component, source A, 
but  also for  the second dust component, source B, represented by a ``hole'' at 
$r \simeq 4\farcs5$  in the right panel of Figure~\ref{poldist}. Those diagrams are 
consistent with Figure \ref{fir5pol}, where the polarization fraction increases with distance
from the peak of emission, but there is a lack of overall polarized emission 
toward source B.

The depolarization observed at higher values of Stokes I seems to be part of a 
global effect observed at  different wavelengths  \citep{Goodman95,Lazarian97}. 
In the mm/submm range, this phenomenon was also observed in the BIMA data 
published  by LCGR02, as well as far-infrared observations with single dishes 
\citep{Schleuning98,Matthews01} . 
The anti-correlation between Stokes I and 
polarization fraction can be caused by different mechanisms. On one hand, it may be 
the result of changes in the grain structure at higher densities. Those changes may 
be responsible for a decrease in the efficiency of dust grain alignment with respect to the 
local magnetic field \citep{Lazarian07b}.  In the case of FIR 5B, the embedded 
source may be in a very early stage of formation, prior even to collapse (since no 
clear evidence of star-forming signatures like outflows has been assigned to it). In 
this case, the lack of internal infrared radiation could provide no radiative torque to 
the dust grains and, therefore, no polarized flux is observed. Another explanation 
for this effect could be a twisted magnetic field or the superposition of distinct
field directions along the LOS resulting in a reduction of the net polarization degree 
\citep{Matthews01}. Observations at higher angular resolution 
would be necessary to resolve the small scale structure.

\subsection{Magnetic field properties}
\label{morf}

\begin{figure}
   \centering
   \includegraphics[angle=-90,width=9cm]{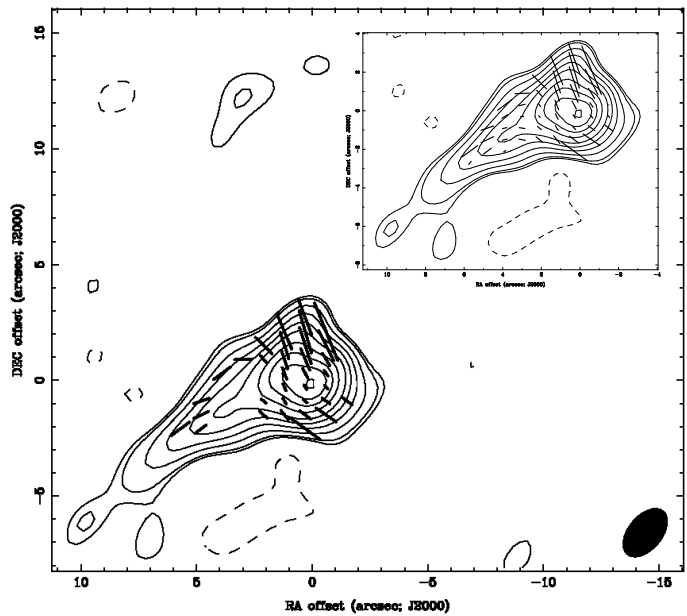}
      \caption{Plane-of-sky field geometry for NGC 2024 FIR 5. Contours, beam and
     vector scale are the same than in Figure \ref{fir5pol}. 
      Vectors are plotted at 2$\sigma_{\mathrm{P}}$ level and relaxed to 
      1$\sigma_{\mathrm{P}}$ in the upper
      right corner.}
         \label{field}
   \end{figure}

In section \ref{intro} we briefly introduce the physical mechanisms associated
with the alignment of dust grains with respect to the magnetic field lines. Although some
works propose that grain alignment could be independent of magnetic fields (e. g.: mechanical 
alignment by particle flux, \citet{Gold52}), it has not been proven yet observationally. Dust grains
are believed to have at least a small fraction of atoms containing magnetic momentum in their 
compositions, so some interaction with the ambient magnetic field is expected. 
The most stable energy state is achieved when the grain longest axis rotates perpendicularly
to the field lines. Consequently, dust emission polarization vectors as observed in submillimeter polarimetry 
have to be rotated by 90$\degr$ in order to be parallel to the plane-of-sky (POS) component of the magnetic field. 
The LOS field component adds no information to the 2D polarization map because the 
spinning dust grains produce zero polarization flux.
If a strong LOS component is expected, a decrease in net polarization flux is observed, and alternative
techniques must be used to measure it (e.g., Zeeman effect observations: \citet{Troland82,Crutcher93}). 
Therefore, the polarization map of Figure \ref{fir5pol}, when rotated by 90$\degr$, traces the projection of the 3D 
magnetic field
morphology on the plane-of-sky (see Figure~\ref{field}). For FIR 5A, the field geometry is described by curved lines 
centered on the protostellar core. 
Toward the elongated emission 
associated with FIR 5B, the field lines are parallel to the core's major axis, 
implying a 90$\degr$ change in the direction with respect to the FIR 5A mean direction.
By relaxing the signal-to-noise level
down to 1--$\sigma$,  one can see that this change in the magnetic field direction 
is not abrupt, and an hourglass morphology can be roughly derived for the main 
component (Figure \ref{field}, upper right box). 
Several theoretical works have performed 3D simulations of collapsing magnetized clouds. 
They all agree that the POS projection of the magnetic field morphology in those class of objects is 
a hourglass shape \citep{Ostriker01,Goncalves08}. Our results, and many others (e.g. \citet{Girart06,Rao09}) provide 
observational support to these models.

So far, the CF relation developed by \citet{CF53} is still the most
straightforward method to estimate the plane-of-sky component of the magnetic field.
Assuming energy equipartition between kinetic and perturbed magnetic energies as
\begin{equation}
\frac{1}{2}\rho\delta V^{2}_{LOS} \simeq \frac{1}{8\pi}\delta B^{2},
\label{equip}
\end{equation}
(where $\delta V_{LOS}$ is the observational rms velocity along the line-of-sight
and $\rho$ is the average  density), this method compares the fraction of uniform
to random components of  the field under effects of Alfv\'enic perturbations
($\delta v \propto \delta B\sqrt{\rho}$) taking into account  isotropic velocity
dispersions. 
The CF formula uses the dispersion of position of polarization angles and
molecular line widths as observational inputs for the gas motions in the core.
However, recent works showed that this approximation overestimates the  magnetic
field for coarser resolutions \citep{Heitsch01,Ostriker01}.  These authors
constrained the application of this method to data sets with relatively low dispersion 
of position angles ($\Delta\theta \leq 25\degr$), which means strong-field cases. By
statistical studies of magnetic turbulent clouds, these authors showed that the CF
formula is  accurate only if this condition is applied. Using the small angle
approximation  $\delta\phi \approx \delta B/B_{uniform}$, the CF formula can be
stated as follows:
\begin{equation}
B_{uniform} = \xi \sqrt{4\pi\rho}\frac{\delta V_{LOS}}{\delta \phi},
\label{CF}
\end{equation}
where $\delta\phi$ is the angle dispersion.
The correction factor $\xi$ ($\simeq 0.5$) arises from the previously mentioned strong field conditions  
to which this case applies \citep{Ostriker01}.
    
Unlike LCGR02, we opted for not applying any geometric model to the observed field
due to the low statistics of our data set. 
The observed dispersion in our data (main component in the histogram of Figure \ref{hist}) is 
12.2$\degr$. 
According to $\delta\phi_{obs} = (\delta\phi^{2}_{int} + \sigma_{\theta}^{2})^{1/2}$, the
observed dispersion depends on the intrinsic dispersion $\delta\phi_{int}$ plus
the measurement uncertainty of the position of the polarization angles
$\sigma_{\theta}$, resulting from the contributions of both effects. 
Since that no geometric models were used to remove the systematic field structure, 
changes on the large-scale field directions are included in the intrinsic dispersion, together with 
turbulent fields due to Alfv\'enic motions.
In our data set, the
position angle uncertainties average 
to 7.52$\degr$, which gives us an intrinsic dispersion of 9.61$\degr$.
Some extra observational parameters are needed to compute the magnetic field
strength with equation \ref{CF}. The average density and the rms
velocity in FIR 5 can be obtained from previous works. \citet{Empre09} modeled
the morphology of NGC 2024 based on APEX observations of CO isotopologues. The
various line profiles obtained for different transitions 
are consistent with a complex  structure composed by a Photo
Dominated Region (PDR) foreground to the molecular gas where the far  infrared
cores are found. In their models, the dense molecular cloud must be warm (75 K) and dense 
($9 \times 10^{5}$ cm$^{-3}$) to reproduce
the velocity gradients observed for distinct cloud components. These results
agree with the previous work of \citet{Mangum99}, based on  formaldehyde
observations. These authors derived a kinetic temperature of $T_{K} >
40$ K for FIR 3-7 and estimate densities at the same order of magnitude
($n_{\mathrm{H_{2}}} \approx 2 \times 10^{6}$ cm$^{-3}$). We adopt
$n_{\mathrm{H_{2}}} = 1.5 \times  10^{6}$ cm$^{-3}$ as an average value for the
density. For the velocity dispersion, we adopt $\delta V_{LOS}$ of 0.87 $\pm$ 0.03 \kms,
which is the value derived by \citet{Mangum99} from the formaldehyde observations.
This molecule is a good tracer of dense gas, and for the single-dish data of
\citet{Mangum99}, it traces the gas kinetic temperature in a scale of $\sim 8000$ AU,  hence
it is well correlated to the turbulent motions of the core.
Finally,  applying those inputs to the equation
\ref{CF}, together with the $\delta\phi_{int}$ previously obtained, we estimate that
the POS magnetic field strength is 2.2 mG, which is in good agreement with the 
value estimated in LCGR02. The uncertainty in the magnetic field strength 
is determined mainly by the error in the volume density $n$, which is a factor of $\sim$ 2
due to the distinct assumptions on the cloud temperature. This factor 
implies an uncertainty of 40\% for the derived field strength.
As mentioned earlier, the dispersion used as input in equation \ref{CF}
carries the combined effects of changes on the large-scale field directions plus 
turbulent motions. 
In this case, the derived field strength is only a lower limit since the angle dispersion
is not generated purely by Alfv\'enic motions. 
On the other hand, beam averaging and line-of-sight effects due to field 
twisting of multiple gas components usually underestimates the real value 
of the turbulent component, and the estimated 
field strength in this case would be an upper limit. So, we can assume that both effects cancel out and 2.2 mG is a fair estimation for the POS field strength.
 
The mass-to-flux ratio gives information on
whether the magnetic field can support the cloud against the gravitational collapse and, therefore, 
it provides clues about the evolutionary state of the source.  
Specifically, this quantity compares the pressure produced by an amount of mass M in a magnetic
tube of flux $\Phi$. A critical value, reached when the magnetic pressure is no longer able to support the 
gravitational pulling, is given by $(2\pi\sqrt{G})^{-1}$ \citep{Nakano78}. 
Observationally, this parameter is defined by \citep{Crutcher99}:
\begin{equation}
\lambda = \frac{(M/\Phi)_{observed}}{(M/\Phi)_{critical}} = (m{\mathrm{N(H_{2}})}A/BA) \times (2\pi\sqrt{G}) = 
7.6 \times 10^{-21} \frac{\mathrm{N(H_{2}})}{B},
\label{mtb}
\end{equation} 
where $(M/\Phi)_{critical}$ is the mass-to-flux ratio of an uniform disk where
gravity is balanced by magnetic pressure, $m = 2.8m_{H}$ allowing for He, $A$ 
is the cloud area covered by observations, 
N(H$_{2}$) is the column density and $B$ is the magnetic field strength.  
Applying the POS magnetic field strength obtained in the previous paragraph, 
$B = 2.2$~mG, and the column densities derived in section \ref{prop_dust}, 
we estimate a mass-to-flux ratio for FIR 5A of 1.6 (for $T = 60$ K),
which corresponds to a core in a supercritical
stage. In any case, those calculations are restricted to the dust envelope,
without taking into account the mass contribution of the embedded protostar.
We consider that the derived mass-to-flux values are only a lower limit for this
quantity and therefore it is in agreement  with the observed star-forming
signatures.   

We can also derive the ratio between turbulent and magnetic energies. 
From the autocorrelation function of the polarization position angles, it is possible to measure how 
the dispersion of PA's varies with respect to the distinct length scales within the cloud. This function
provides an indirect calculation of the turbulent to magnetic energy as 
\citep{Hildebrand09}:
\begin{equation}
\label{bturb}
\beta_{turb} \approx 3.6 \times 10^{-3} \left ( \frac{\delta\phi}{1\degr} \right)^{2}
\end{equation} 
For the angular dispersion obtained from our sample, $\delta\phi_{int} = 9.61$, we
compute the turbulent to magnetic energy ratio as $\beta_{turb} = 0.33$.  
This value agrees with the ratio estimated in LCGR02,
which reinforces that the turbulent motions are magnetically dominated. The turbulent-to-magnetic energy 
ratio found for FIR 5 is consistent with what was measured for other low-mass protostellar cores like
NGC~1333 IRAS~4A and IRAS $16293-2422$ \citep{Girart06,Rao09}.

\subsection{Magnetic field around FIR 5A: gravitational pulling or H{\sc ii} compression?}

In this section, we try to elucidate which mechanism is mainly responsible
for the observed curved magnetic field morphology in FIR 5. One possibility is that
the gravitational pulling overcomes the local magnetic support and drags the 
ionized  material toward the center, warping the field lines in such a way that they 
assume an hourglass morphology. This is consistent with the previous result that  
the protostellar core is in the supercritical regime. However, if this is the case then 
only the hourglass component west of FIR 5A is observed. 
The lack of detected vectors east of FIR 5A could be due to the overlap in the line of 
sight of the dust polarization associated with both FIR 5A and FIR 5B cores. 
The magnetic field direction associated with FIR 5B is perpendicular to the 
FIR 5A main direction. Alternatively if the two cores are connected, then it could be due to an abrupt 
change in the magnetic field direction. In both cases, the net polarization flux 
is expected to decrease significantly.
Another possibility is that the grain alignment efficiency associated with source B is smaller. Indeed,  
Figure \ref{poldist}b (right panel) 
show that the second polarization ``hole'' matches quite well to the position of FIR 5B.  
Of course, the cause could also be a combination of these possibilities. 

If the tension generated by the magnetic field curvature is produced by the gravitational collapse, then we 
can make a rough estimation of the mass required to produce the observed curvature. This magnetic force 
can be expressed as $B^{2}/R$, where R is 
the radius of curvature of the field lines. According to the equations derived by 
\citet{Schleuning98}, we have
\begin{equation}
\left[ \frac{M}{100 \mathrm{M_{\odot}}} \right] = \left[ \frac{B}{1 \mathrm{mG}}\right]^{2}  
\left[ \frac{D}{0.1 \mathrm{pc}}\right]^{2}  
\left[ \frac{R}{0.5 \mathrm{pc}} \right]^{-1} 
 \left[ \frac{n(\mathrm{H_{2}})}{10^{5}\mathrm{cm^{-3}}} \right]^{-1}
\label{mass}
\end{equation}
where $D$ is the distance of the field lines from the protostar. 
At $D=1\farcs9$ (789 AU) the field lines have a radius of curvature R of $17''$
(7055 AU).  At the selected radius of curvature, the estimation of magnetic tension 
force is  $\sim 10^{-23}$ dyne cm$^{-3}$. Applying these values to equation \ref{mass}, 
we find that the mass inside the radius of $1\farcs9$ is  $\simeq2.3$~M$_{\odot}$. 
Although this value is almost a factor of two higher than the mass 
estimation done for FIR 5A in section \ref{prop_dust}, it is within the same order of magnitude 
of the first estimation, even with the large uncertainties in the assumptions of $D$ and the radius of
curvature $R$. This method is an alternative approximation to test if gravitational pulling is the
responsible for the magnetic field curvature.

\begin{figure}
   \centering
   \includegraphics[width=12cm]{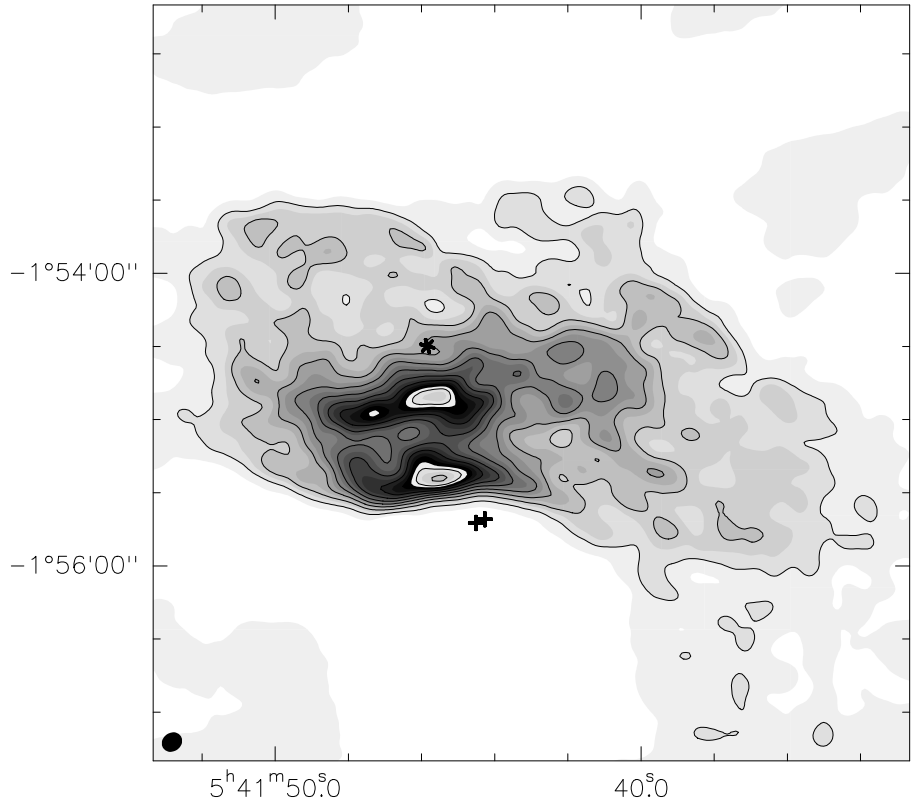}
   \caption{VLA 6 cm emission from the H{\sc ii} region in NGC 2024 (data from the
      VLA Archive). Grey scale-filled contours are 0.5,  1.5, 2, 3, 4, 5, 6, 7,
      9, 11, 13.1$\sigma$ (1$\sigma \simeq 9$ \mJybm). The beam size of 
      $8.6^{\prime\prime} \times 7.5^{\prime\prime}$ and PA of 37$\degr$ 
      is shown in the lower left corner.
      Crosses indicate the location of FIR 5A and FIR 5B sources. The star
      indicates the position of the ionization source.}
         \label{regHii}
   \end{figure}

Given the situation of the FIR 5 core, the external agents may also interfere in the protostellar physical  
environment. Previous observations proved that the molecular ridge
and the chain cores in NGC 2024 are located at the far side of the H{\sc ii} region
\citep{Barnes89,Schulz91,Chandler96,Crutcher99}. The distribution of molecular
and ionized gas proposed by \citet{Matthews02}  for NGC 2024 (Figure 8 in their
paper) has the western portion of the ionization front expanding toward the
background  molecular ridge and stretching the magnetic field lines  around the 
ridge of dense cores. At large scales ($\sim 0.5$~pc), this morphology is corroborated 
not only by the LOS field obtained from the CN Zeeman observations \citep{Crutcher99}
but also by the POS field from the single-dish dust polarization data. 
At smaller scales ($\sim 0.02$ pc), this could have 
an effect of  compressing the magnetic field lines, bending them toward the 
east, as observed around the FIR 5 core.
In order to check if the radiation pressure can be large enough to compress the
molecular gas,  we have studied the distribution of the ionized gas in NGC
2024. For this purpose, we accessed the NRAO Data Archive System to search for
centimeter emission that could reproduce this morphology. We found an
extended emission in 6 cm related to the H{\sc ii} region produced by the
star IRS 2b.  Figure~\ref{regHii} shows that the hot gas has an extended component 
to the west and is roughly flattened to the south (although slightly curved to the 
southwest). FIR 5A and FIR 5B, indicated as crosses in Figure~\ref{regHii}, lie in the border 
of the H{\sc ii} region. 
The bright southern pattern near FIR~5 could trace the compressed ionized gas resulting 
from the interaction between hot/diffuse and cold/denser components. The radiation 
pressure produced by 
the illuminating star ($P_{\mathrm{rad}}$) can be  calculated by $\frac{L}{cA}$, 
where $L$ is the luminosity of the ionization source, $c$ is the speed of light and 
$A$ is the area of the expanding shell. \cite{Bik03}  found that the spectral type of 
IRS 2b is in the range O8 V--B2 V, which is consistent with the intensities of radio
continuum  and recombination lines observed in the H{\sc ii} region
\citep{Kruegel82,Barnes89}. Therefore, we can assume $L = 10^{5.2}  L_{\odot}$,
which is representative of such a spectral type. A first estimation for the radius of
the H{\sc ii} region was done by \citet{Schraml69} through low
resolution ($\sim 2^{\prime}$) radio observations of NGC 2024. These authors
measured a radius of 0.2 pc ($\simeq 41 \times 10^{3}$ AU) inferred from the 
observed emission. However, from Figure~\ref{regHii}, the radius of the centimeter emission 
can be fairly estimated in $\sim 1^{\prime}$, which is approximately the distance 
between FIR~5  and IRS~2b. 
As a result, the radiation pressure $P_{\mathrm{rad}}$ is calculated as 1.16
$\times 10^{-8}$ dyne cm$^{-2}$. The ionization pressure ($N_{e} \times T_{e} \times k$) 
also accounts for the energy produced by the PDR. We assume an electron 
density of 5.94 $\times 10^{3}$ cm$^{-3}$ as derived from the emission parameters 
of the centimeter VLA map.
The recombination line studies of \citet{Reifenstein70} provide an electron 
temperature of 7200 K for this H{\sc ii} region. To be conservative, we adopt a range 
of 7200--15000 K for $T_{e}$. With these parameters,
the ionization pressure is estimated to range between $5.9\times 10^{-9}$ and 
$1.2\times 10^{-8}$~dyne~cm$^{-2}$.  On the other hand, the magnetic pressure 
is defined by $P_{mag} = \frac{B^{2}}{8\pi}$, where $B$ is the total magnetic field 
strength. Since our SMA maps provide a two-dimensional picture of the total field, we are able to 
calculate only a lower limit for the magnetic pressure. Therefore, applying the previous equation for the field strength obtained in section \ref{morf}, we have   
$ P_{mag} \geq 1.96\times10^{-7}$~dyne~cm$^{-2}$. This value is at least one order of 
magnitude higher than the radiation and ionization pressures. 
Even if we add the thermal pressure to the calculations ($P_{ther} \simeq 1.4 
\times 10^{-9}$ dyne cm$^{-2}$, \citet{Vallee87}), the energy injected by
the ionization front is still lower than the magnetic force. Therefore the expanding 
H{\sc ii} region is not enough to compress the magnetic field lines into the 
observed geometrical configuration, and the bending is produced by the gravitational 
pulling.

\subsection{Multiple Outflows}
\label{outflows}

Previous works reported that FIR 5 has an associated (redshifted) unipolar and highly 
collimated outflow with a mass of $\sim 4$~M$_{\odot}$ and density of 
$\sim 10^{2}$~cm$^{-3}$ \citep{Sanders85,Richer92,Chandler96}. However, a rather 
complex morphology was proposed by \citet{Chernin96} as indicated by their 
interferometric (BIMA) and single dish (NRAO 12 m telescope) combined maps of 
CO ($1 \rightarrow 0$). In those, in addition to the unipolar lobe associated to FIR~5, 
there are two other redshifted features along the North-South direction and 
practically parallel to the FIR 5 molecular outflow, but neither of them associated 
with it. These two components were named $ns1$ and $ns2$ and are detected 
at \vlsr velocities between 15 and 25~\kms. $ns2$ is associated with FIR~6.
They also identified a blueshifted feature, $ew1$, extending east of FIR~6.
\citet{Chernin96} proposed that the brightest outflow component apparently 
powered by FIR 5A has a layered velocity structure. Their lower resolution 
combined molecular maps ($\sim 4\farcs5$) are dominated by an unipolar red 
lobe composed by two parallel outflows at lower velocities ($\sim$ 20 \kms ~in 
their Figure 1)  which merge into only one at high velocities . This redder emission, 
referred as $ns1$ in their  paper, is more collimated than the lower velocity 
components and arises $10''$ west of FIR  5: A. The author suggests that the 
$ns1$ outflow is widened by jet-wandering or internal shocks \citep{Chernin95} 
and is powered by a deeply embedded and undetected source rather than FIR 5A 
due to its misalignment with it.

In this work, we offer a different interpretation for this scenario. The SMA CO $(3\rightarrow2)$ 
maps have an angular resolution a factor of 2 higher than the combined maps of \citet{Chernin96}.  
Contrary to the suggestion by \citet{Chernin96}, our maps show that 
this outflow is clearly powered by FIR 5A instead of by a faint, undetected low-mass star. 
The overall morphology described by \citet{Chernin96} is
also observed in our maps. The main difference is that we detect high velocity
gas which is offset by $10''$ west of FIR 5A, coinciding in position with the previously
undetected FIR 5-sw dust condensation.Then, two interpretations can be derived from
those features. Firstly, it is possible that all components are part of a single but velocity-layered outflow,
the two low velocity lobes tracing the cavity where the highest velocity
outflow is located. The presence of the high
velocity lobe not only at the center of the cavity but also displaced from it could suggest that
the outflow is precessing. Alternatively, the presence of the FIR 5-sw 
dust source associated with the western red shifted lobe, and in particular at high 
outflow velocities, suggest that this lobe could be an independent molecular outflow. 
As in the case of FIR 5A, this outflow would be also an unipolar outflow. 

\begin{figure}
 \centering
 \epsscale{0.5}
 \plotone{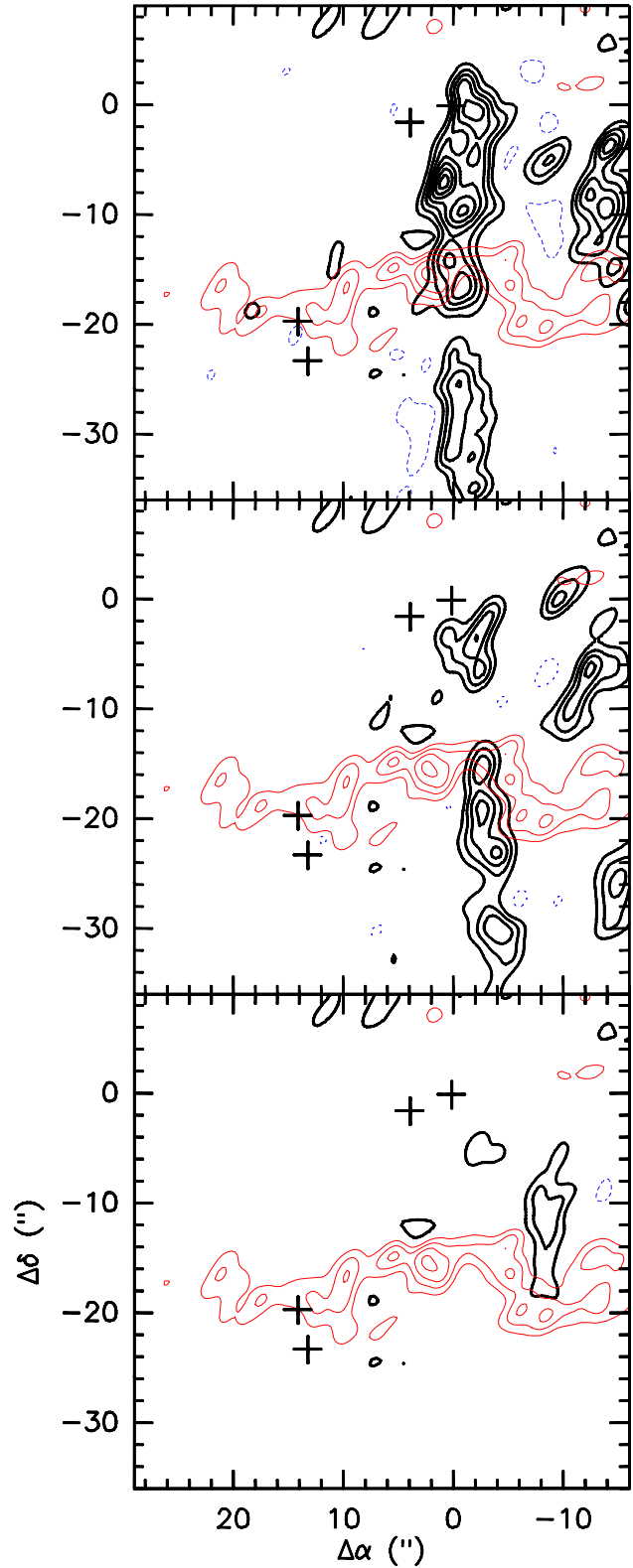}
\caption{Superposition of contours from 15.6, 19.2 and 29.7 \kms velocity channels 
(black contours) over 7.2 \kms channel (red contours). Intensity contours are
-3, 3, 4, 5, 7, 9, 12 $\sigma$ (1 $\sigma \approx 0.57$ \mJybm). Source positions are 
maked as crosses.}
\label{CO32superpos}
\end{figure}

\begin{figure}
\centering
\includegraphics[angle=-90,width=13.5cm]{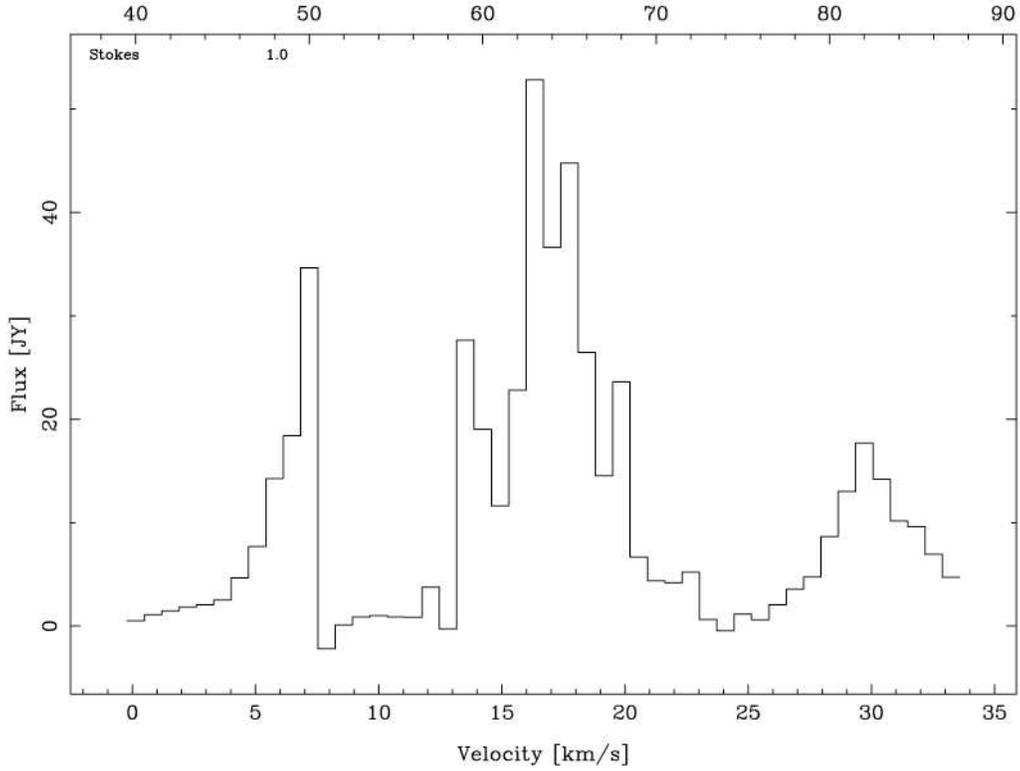}
\caption{Spectrum of the interacting zone between the EW outflow, powered by FIR 6, and 
the high velocity feature apparently powered by FIR 5-sw. The spectrum was obtained for a 
velocity range of 0 to 33.2 \kms. The three peaks correspond to the emission from the EW outflow 
(the blue shifted peak at $\sim$ 7 \kms), the main lobe powered by FIR 5A 
at $\sim$ 18 \kms and the high velocity lobe
arising from FIR 5-sw at $\sim$ 30 \kms.}
\label{espectro}
\end{figure}

Our CO $(3\rightarrow2)$ maps seem to indicate a possible interaction between the different outflows.
In the blue lobe of Figure \ref{CO32}, there is extended emission centered in the dust condensation
FIR 6n (using the LCGR02 nomenclature) with an EW orientation. The emission is
associated with an unipolar outflow detected from $\sim$ 1.0 to 7.8 \kms~and is characterized by a
wandering/wiggling morphology. The outflow suffers structural changes in its shape, represented by drastic 
rotations in PA. From being powered initially toward the NW direction, it assumes an almost horizontal 
distribution (PA = 94\degr) which corresponds to its brightest emission. Then, another structural change 
seen at 7.8 \kms~results in an $u-$like shape. Both the bending 
from NW to the EW direction and the $u-$like 
structure coincide spatially with the projection of the red-lobe NS outflows powered by FIR 5A and FIR 5: 
sw. Figure \ref{CO32superpos} shows contours of velocity channels
15.6, 19.2 and 29.7 \kms~(black contours) overlaid with the  7.2 \kms~channel (red contours) 
in the upper, middle and bottom panels, respectively. 
The variations in PA of the EW outflow may be somehow due to the
interaction with the main outflow powered by FIR 5A and the high velocity emission from FIR 5-sw. 
Global inspection of these three
panels tentatively leads to the hypothesis of a shock interaction between distinct outflows in such a way
that the gas structure is modified. 
The spectrum exhibited in Figure \ref{espectro} was obtained for the supposedly interacting zone of the 
panels of Figure \ref{CO32superpos}. A box of $-6'' < \Delta\alpha < -18''$ and $-9'' < \Delta\delta < 
-24''$ was used to select a region where all three outflows components (EW, main and high velocity 
features) contribute to the emission. The lack of CO emission seen at $\sim$ 25 \kms~is probably due
to the cavity previously mentioned and corresponds to the velocity interval between low and high red
velocity components. The blue emission is related to the EW outflow.

\subsection{Unipolar molecular outflow}

Previous observations of NGC 2024 fail to detect a blue counterpart 
for the bright outflow powered by FIR 5A \citep{Richer92,Chernin96}. However, 
other studies claim a bipolarity for this outflow \citep{Sanders85,Barnes89}. 
In all cases, the red shift lobe which is ejected toward the south has a brighter, 
more extended and collimated emission than the putative blue counterpart. 
These works all identify this feature as the main component of this outflow, making 
the morphology of the blue lobe, if it really exists, of unclear nature.

This asymmetrical pattern of the outflow powered by FIR 5A
could be explained by the cloud morphology proposed by \citet{Matthews02}, 
illustrated in Figures 6 and 8 in their paper. They indicate how the
NGC 2024 H{\sc ii} region could be seen from the west and north directions, 
respectively. In this scenario, the dust
cores appear in the dense molecular cloud behind the H{\sc ii}  expanding front (using the 
line-of-sight as reference), with FIR 5 projected right below the interface zone. 
Consequently, the bipolar molecular outflow ejected from this core will have the
blue molecular component destroyed by the UV photons produced in the H{\sc ii} region,
since it points right toward it, but the red lobe would remain intact.  
Despite the fact that
\citet{Barnes89} did not know accurately which core is the driving source of the supposedly blue outflow, 
these authors found that the total luminosity of the nebula is comparable to the flow energy. Therefore, some 
interaction between both could be expected.

\section{Conclusions}

In this paper, we report SMA polarization observations of the intermediate-mass
protostellar core NGC 2024 FIR 5.  Data acquisition was done using the
polarimetric capabilities of the SMA combined with wide spectral window
receivers. The polarized flux appears distributed in three
components: two of them around the peak of total intensity (Stokes I) and another
component arising from the elongated portion of the  core. The overall
polarization portion resembles a partial hourglass morphology due to a possible
ambipolar diffusion  phenomenon taking place in the core.

The magnetic field strength was  estimated in 2.2 mG.
The estimates of turbulent-to-magnetic energy and mass-to-flux ratio are consistent with a
supercritical highly magnetized core. In previous works, magnetized collapsing 
cores were also observed in high-mass protostars. In general, ambipolar diffusion 
seems to affect core evolution globally, independent of the mass range.

The absence of a symmetrical field morphology gives rise to different interpretations
for the field structure in the core.  The dust cores in NGC 2024 may be affected
by an expanding ionization front compressing the molecular gas. It could be
perturbing the field structure at smaller scales. The bended lines observed in
our SMA maps could be the  consequence of the radiation pressure of the hot
component. Previous VLA 6 cm observations trace the foreground H{\sc ii} region
as an extended emission produced by the O2--B2 ionization source IRS 2b. However,
our estimations of radiation pressure due to  the expanding shell does not
overcome the magnetic pressure generated by the field lines. So, the asymmetrical magnetic field is
more likely due to depolarization effects arising in the position of the previously unresolved
FIR 5B source. 

A complex outflow morphology was observed toward FIR 5. Several 
collimated features were detected toward FIR 5, FIR 6 and FIR 5-sw. We speculate
about a possible flow interaction between distinct components. It could explain the 
structural changes observed in some outflows. 
The brighter emission powered by FIR 5A has a clumpy structure and arises 
highly collimated in a NS orientation. The absence of a blue lobe counterpart can be
attributed to the expanding H{\sc ii} region to the north of the core. The UV
radiation field may be responsible for dissociating the molecular structure of the
outflow, destroying this component.

\acknowledgments

FOA would like to thank all the staff at the SMA in Hawaii and Cambridge, MA. FOA and JMG are supported by the 
MICINN AYA2008-06189-C03 and the AGAUR 2009SGR1172 grants.

{\it Facilities:} \facility{SMA (polarimetry)}.

\clearpage

\clearpage

\clearpage

\clearpage

\clearpage

\clearpage

\clearpage

\clearpage

\begin{deluxetable}{ccccccc}
\tabletypesize{\scriptsize}
\tablecaption{Parameters of the continuum and line observations\label{table_cont}}       
\tablewidth{0pt}
\tablehead{
\colhead{Observations} &
\colhead{Rest frequency}  & 
\colhead{HPBW}  & 
\colhead{PA\,\tablenotemark{a}} & 
\colhead{Spectral resolution} &
\colhead{Peak of emission} & 
\colhead{$rms$ noise}  \\
\colhead{} & \colhead{(GHz)} & \colhead{(arcsec)}  & \colhead{($\degr$)} & \colhead{\kms} & 
\colhead{(\Jybm)}  & \colhead{(\Jybm)} \\
}
\startdata
 Continuum & 345.8000 & $2.45 \times 1.48$ & -39.8 & -- &1.19 & 0.019\,\tablenotemark{b} \\                       
 CO (3$\rightarrow$2) & 345.7960 & 2.87 $\times$ 1.67 & -37.6 & 0.7 & 9.5 & 0.57 \\
\enddata
\tablenotetext{a}{Position angles are measured from North to East.}
\tablenotetext{b}{The $rms$ noise of Stokes I emission, obtained with a robust weight of 0.5}
\end{deluxetable}

\clearpage

\begin{deluxetable}{cccccccc}
\tabletypesize{\scriptsize}
\tablecolumns{8} 
\tablecaption{FIR 5: main component\tablenotemark{a}\label{fir5main_table}}       
\tablehead{
\colhead{Dust} &
\colhead{$\alpha_{2000}$\tablenotemark{b}} &
\colhead{$\delta_{2000}$\tablenotemark{b}} &
\colhead{Peak\tablenotemark{b}} &
\colhead{Total\tablenotemark{c}} &
\colhead{FWHM} &
\colhead{Deconvolved\tablenotemark{c}} &
\colhead{Deconvolved\tablenotemark{c}} \\
\colhead{condensation} & \colhead{} & \colhead{}  & 
\colhead{of intensity} & 
\colhead{flux} & 
\colhead{Gaussian fit} & \colhead{size} & \colhead{PA} \\
\colhead{} & \colhead{} & \colhead{} &  \colhead{(Jy beam$^{-1}$)} &
\colhead{(Jy)} & \colhead{(arcsec)} & \colhead{(arcsec)} &  \colhead{($\degr$)}  
}
\startdata
FIR 5A & 05 41 44.258 & $-01$ 55 40.94 & 1.16(2) & 2.44(6) &
	2.8 $\times$ 2.4 & 2.31(6)$\times$1.58(8) & 47(5) \\
FIR 5B & 05 41 44.510 & $-01$ 55 42.35 & 0.40(2) & 1.26(8) 
	& 3.6 $\times$ 2.8 & 3.1(2)$\times$2.3(1) & 130(11) \\
\enddata
\tablecomments{Units of right ascension are hours, minutes and seconds, and units of
declination are degrees, arcminutes, and arcseconds.}
\tablenotetext{a}{Fit error of the last digit in parenthesis.}
\tablenotetext{b}{Estimated using Miriad's ``MAXFIT'' task.}
\tablenotetext{c}{Values derived with AIPS's ``IMFIT'' task.}
\end{deluxetable}

\clearpage

\begin{deluxetable}{ccccc}
\tablecolumns{5} 
\tablecaption{Sub-millimeter dust condensations\label{condens}}       
\tablewidth{0pt}
\tablehead{
\colhead{Dust} &
\colhead{$\alpha_{2000}$\,\tablenotemark{b}} &
\colhead{$\delta_{2000}$\,\tablenotemark{b}} &
\colhead{Peak} &
\colhead{Flux}  \\
\colhead{condensation} & \colhead{} & \colhead{}  & \colhead{of intensity\,\tablenotemark{b}} & \colhead{density} \\
\colhead{} & \colhead{} & \colhead{}  & \colhead{(mJy beam$^{-1}$)} & \colhead{(mJy)}
}
\startdata
FIR 5-sw & 05 41 43.667 & -01 55 49.05 & 77(18) & 81(21) \\
FIR 5-ne & 05 41 45.043 & -01 55 31.60 & 131(18) & 202(30) \\
FIR 6n\,\tablenotemark{a} & 05 41 45.193 & -01 56 00.50 & 70(18) & 40(15)  \\
FIR 6c\,\tablenotemark{a} & 05 41 45.134 & -01 56 04.01 & 124(18) & 116(22) \\
\enddata
\tablenotetext{a}{According to \citet{Lai02} numbering.}
\tablenotetext{b}{Estimated using Miriad's ``MAXFIT'' task.}
\end{deluxetable}

\clearpage

\setlength{\tabcolsep}{0.035in}
\begin{deluxetable}{cccccccc}
\tablecolumns{8} 
\tablecaption{SMA polarization data from NGC 2024 FIR 5\label{table}}
\tablewidth{0pt}
\tablehead{
\colhead{$\Delta$ RA\,\tablenotemark{a}}  & 
\colhead{$\Delta$ Dec} & 
\colhead{P} & 
\colhead{$\epsilon$P}  & 
\colhead{$\sigma_{P}$\tablenotemark{b}} & 
\colhead{I$_{P}$\tablenotemark{c}} & 
\colhead{$\theta$\tablenotemark{d}}  & 
\colhead{$\epsilon\theta$}   \\
\colhead{(arcsec)} & 
\colhead{(arcsec)} & 
\colhead{(\%)} & 
\colhead{(\%)} &
\colhead{} & 
\colhead{(\Jybm)} & 
\colhead{($\degr$)} & 
\colhead{($\degr$)} 
}
 \startdata
5.4     & -2.1  & 6.00  & 2.0   & 3.00       &      0.018 &     35.723     & 9.194   \\      
0       & -2.1  & 13.4  & 3.1   & 4.32    &      0.027 &  -34.364       & 6.093   \\
5.4     & -1.5  & 9.20  & 2.8   & 3.29    &      0.019 &  27.954        & 8.355   \\ 
4.5     & -1.5  & 3.40  & 1.4   & 2.43    &      0.013 &  24.122                & 12.244  \\
1.8     & -1.5  & 3.60  & 1.6   & 2.25    &      0.013 &  -48.573       & 12.838  \\ 
0.9     & -1.5  & 3.30  & 1.2   & 2.75    &      0.016 &  -47.342       & 10.207  \\ 
0       & -1.5  & 5.20  & 1.1   & 4.73    &      0.027 &  -35.171       & 6.1     \\      
-0.9    & -1.5  & 8.00  & 1.9   & 4.21    &      0.025 &  -34.346       & 6.549   \\
1.8     & -0.9  & 2.00  & 1.0   & 2.00       &      0.012 &    -49.173     & 13.955  \\
0.9     & -0.9  & 1.70  & 0.7   & 2.43    &      0.014 &  -55.723       & 11.734  \\ 
0       & -0.9  & 1.80  & 0.6   & 3.00       &      0.017 &  -39.46        & 9.757   \\ 
-0.9    & -0.9  & 3.40  & 1.0   & 3.40     &      0.019 &    -37.513     & 8.696   \\ 
1.8     & -0.3  & 1.50  & 0.7   & 2.14    &      0.012 &  -60.593       & 13.363  \\ 
0.9     & -0.3  & 2.10  & 0.6   & 3.50     &      0.022 &  -64.888       & 7.45    \\ 
0       & -0.3  & 1.40  & 0.5   & 2.80     &      0.016 &  -52.705       & 9.965   \\ 
-0.9    & -0.3  & 1.60  & 0.8   & 2.00       &      0.012 &  -47.955       & 13.225  \\ 
3.6     &  0.3  & 4.80  & 2.4   & 2.00       &      0.012 &  32.233                & 13.901  \\
1.8     &  0.3  & 2.40  & 0.7   & 3.43    &      0.019 &  -64.96                &  8.479  \\
0.9     &  0.3  & 4.00  & 0.5   & 8.00       &      0.042 &  -67.692       & 3.88    \\
0       &  0.3  & 2.80  & 0.5   & 5.60     &      0.031 &  -58.781       & 5.187   \\ 
-0.9    &  0.3  & 1.70  & 0.7   & 2.43    &      0.013 &  -56.207       & 12.44   \\ 
2.7     &  0.9  & 3.90  & 2.1   & 1.86    &      0.011 &  -10.944       & 15.247  \\ 
1.8     &  0.9  & 3.90  & 0.9   & 4.33    &      0.023 &  -62.435       & 6.924   \\ 
0.9     &  0.9  & 6.70  & 0.6   & 11.2   &      0.061 &  -70.852       & 2.667   \\ 
0       &  0.9  & 6.30  & 0.7   & 9.00       &      0.051 &  -63.767       & 3.165   \\ 
-0.9    &  0.9  & 3.30  & 1.0   & 3.30     &      0.018 &    -58.229     & 8.806   \\ 
1.8     &  1.5  & 6.80  & 1.7   & 4.00       &      0.023 &  -57.537       & 7.051   \\ 
0.9     &  1.5  & 9.40  & 0.9   & 10.4   &      0.061 &  -73.461       & 2.66    \\ 
0       &  1.5  & 11.7  & 1.1   & 10.6   &      0.063 &  -69.462       & 2.569   \\ 
-0.9    &  1.5  & 8.60  & 2.0   & 4.30     &      0.025 &    -61.584     & 6.407   \\ 
0.9     &  2.1  & 11.1  & 1.5   & 7.40     &      0.043 &  -71.543       & 3.729   \\ 
0       &  2.1  & 15.0  & 1.7   & 8.82    &      0.054 &  -72.681       & 2.981   \\
0.9     &  2.7  & 14.2  & 3.1   & 4.58    &      0.028 &  -61.117       & 5.685   \\ 
0       &  2.7  & 12.6  & 2.7   & 4.67    &      0.029 &  -71.458       & 5.554   \\ 

\enddata
\tablenotetext{a}{Offset respect to the peak of total intensity (same for declination).}
\tablenotetext{b}{Signal-to-noise ratio of polarization.} 
\tablenotetext{c}{Polarized intensity $\times 10^{-2}$.}
\tablenotetext{d}{Position angles are measured from North to East.}
\end{deluxetable}


\end{document}